\documentclass{article}
\usepackage{waspaa21,amsmath,graphicx,times}
\usepackage{color}
\usepackage[table]{xcolor}%
\usepackage{booktabs} %
\usepackage{tabularx}
\usepackage{url}
\usepackage[hidelinks,bookmarks=true,hypertexnames=true]{hyperref}  %
\usepackage{multirow} %
\usepackage{cite}
\usepackage[T1]{fontenc}

\newcommand{\around}{{\raise.17ex\hbox{$\scriptstyle\sim$}}}
\newcommand{\hermconj}{\mathsf{H}}
\newcommand{\transpose}{\mathsf{T}}   %

\renewenvironment{thebibliography}[1]{
  \begin{oldthebibliography}{#1}
  \setlength{\itemsep}{0.1em}
  \setlength{\parskip}{0.1em}
}
{
  \end{oldthebibliography}
}

\title{Closing the Gap Between Time-domain Multi-channel Speech Enhancement on Real and Simulation Conditions}

\name{Wangyou Zhang$^{1}$,
      Jing Shi$^{2}$,
      Chenda Li$^{1}$,
      Shinji Watanabe$^{3\dagger}$,
      Yanmin Qian$^{1\dagger}$\vspace{-.5em}\thanks{$^{\dagger}$Yanmin Qian and Shinji Watanabe are the corresponding authors.}}
\address{$^1$ MoE Key Lab of Artificial Intelligence, AI Institute\\
X-LANCE Lab, Department of Computer Science and Engineering\\
Shanghai Jiao Tong University, Shanghai, China\\ \texttt{\{wyz-97, lichenda1996, yanminqian\}@sjtu.edu.cn}\\
         $^2$ Institute of Automation, Chinese Academy of Sciences\quad \texttt{shijing2014@ia.ac.cn}\\
         $^3$ Carnegie Mellon University, USA\quad \texttt{shinjiw@ieee.org}
\vspace{-1.2em}}

\begin{document}

\ninept
\maketitle

\begin{sloppy}

\begin{abstract}
\vspace{-.3em}
  The deep learning based time-domain models, e.g.~Conv-TasNet, have shown great potential in both single-channel and multi-channel speech enhancement. However, many experiments on the time-domain speech enhancement model are done in simulated conditions, and it is not well studied whether the good performance can generalize to real-world scenarios.
  In this paper, we aim to provide an insightful investigation of applying multi-channel Conv-TasNet based speech enhancement to both simulation and real data. Our preliminary experiments show a large performance gap between the two conditions in terms of the ASR performance. Several approaches are applied to close this gap, including the integration of multi-channel Conv-TasNet into the beamforming model with various strategies, and the joint training of speech enhancement and speech recognition models.
  Our experiments on the CHiME-4 corpus show that our proposed approaches can greatly reduce the speech recognition performance discrepancy between simulation and real data, while preserving the strong speech enhancement capability in the frontend.
\end{abstract}

\vspace{-.3em}
\begin{keywords}
multi-channel speech enhancement, time domain, beamforming, automatic speech recognition
\end{keywords}

\vspace{-1.2em}
\section{Introduction}
\label{sec:intro}
\vspace{-0.9em}
With the development of deep learning, speech enhancement (SE), as well as speech separation, has witnessed remarkable advances in both single-channel and multi-channel scenarios~\cite{Conv_TasNet-Luo2019,Divide-Liu2019,DNN_supported-Nakatani2020,Sequential-Wang2021}. Since surprisingly good performance has been achieved in the simulated conditions, more and more researches have drawn their interests in more realistic environments, such as noisy and reverberant speech recorded in various real-world scenarios.

\vspace{-0.05em}
When multiple microphones are available, the capacity of deep learning based speech enhancement models can be further boosted by leveraging the additional spatial information between different channels. A straightforward way is to apply single-channel speech enhancement techniques to the multi-channel speech by extracting the spatial feature as an auxiliary input~\cite{Multi_channel-Chen2018,Neural-Gu2019,Enhancing-Gu2020}.
However, such approaches inevitably introduce artifacts to the enhanced signal, which can be harmful to the downstream automatic speech recognition (ASR) task~\cite{Beam_TasNet-Ochiai2020}, even though the artifacts are imperceptible to human listeners.
Another widely adopted method is known as the neural beamformer~\cite{Neural-Heymann2016, Improved-Erdogan2016}. It usually consists of a mask estimation network for predicting time-frequency masks and a conventional beamformer module such as the minimum variance distortionless response (MVDR)~\cite{Beamforming-VanVeen1988} beamformer. The neural beamformer is favored for its good compatibility with the downstream ASR task, as it explicitly constrains the enhanced output to be distortionless and thus enjoys better generalizability in realistic scenarios.

\vspace{-0.1em}
More recently, the time-domain audio separation network (TasNet)~\cite{TasNet-Luo2018, Conv_TasNet-Luo2019} was proposed for speech separation, and was later extended for denoising~\cite{Improving-Kinoshita2020}. Different from conventional frequency-domain approaches, TasNet directly operates on the input waveform and performs speech enhancement on the learned representation space. It shows very promising results on several benchmarks in both single-channel speech enhancement and separation~\cite{Conv_TasNet-Luo2019,Improving-Kinoshita2020,WHAM-Wichern2019,LibriMix-Cosentino2020}.

\vspace{-0.05em}
While the aforementioned time-domain approaches bring significant performance improvement to speech enhancement, the performance gap between real and simulation conditions is still widely observed~\cite{CHiME4-Vincent2017,Improving-Kinoshita2020,Permutation-Liu2021}.
In this paper, we aim to reduce the gap between time-domain multi-channel speech enhancement on real and simulation conditions, which has not been well studied yet. One interesting direction is the combination of TasNet and neural beamforming. TasNet has strong modeling capability, and MVDR beamforming has the benefit of enhancement without distortion. But how these two methods can benefit from each other in the multi-channel speech enhancement task is not well studied.
Previous work~\cite{Beam_TasNet-Ochiai2020} proposed the Beam-TasNet to estimate the beamformer filter on the output of a multi-channel Conv-TasNet (MC-Conv-TasNet)~\cite{End_to_End-Gu2019,Beam_TasNet-Ochiai2020} for speech separation, which demonstrates superior performance over the vanilla MC-Conv-TasNet and oracle MVDR beamformer.
However, the experiments were conducted only on simulated mixture data, without any background noise. Therefore, the performance and robustness of this approach on realistic data are still unknown.

\begin{figure*}
  \centering
  \centerline{\includegraphics[width=\textwidth]{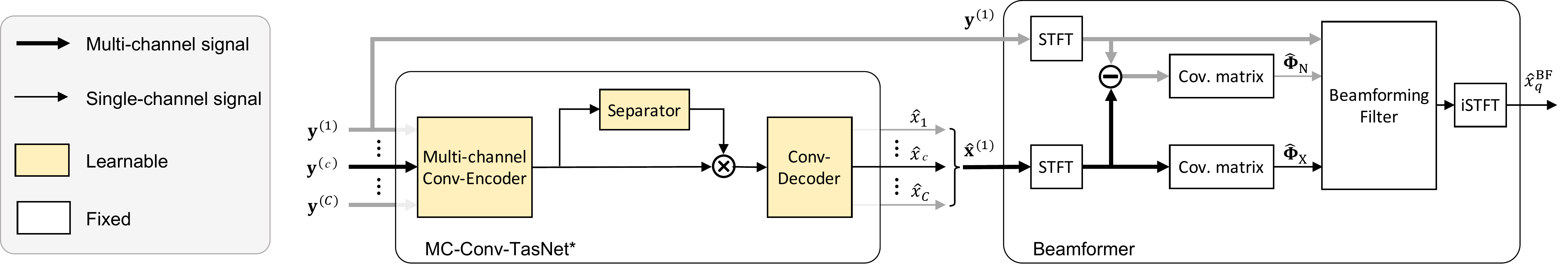}}
  \vspace{-0.5em}
  \caption{Schematic diagram of the Beam-TasNet architecture for multi-channel speech enhancement. Gray lines denote different channel-rotated input signals, which are fed into the MC-Conv-TasNet* module separately to form the multi-channel enhanced signal $\hat{\mathbf{x}}^{(1)}$.}
  \label{fig:beam_tasnet}
  \vspace{-1em}
\end{figure*}
\vspace{-0.05em}
In this work, we show that both MC-Conv-TasNet and Beam-TasNet trained on simulated noisy data can suffer from severe performance degradation on the real data.
To alleviate such degradation, we propose two training schemes to improve the performance and robustness of MC-Conv-TasNet and Beam-TasNet:
(1) Exploring different integration approaches in the Beam-TasNet framework; (2) Joint training of MC-Conv-TasNet and ASR models.
We evaluate different methods on the CHiME-4~\cite{CHiME4-Vincent2017} corpus, which consists of real and simulation data for both training and testing, allowing us to verify the performance gap in different conditions.
For real data, since it is hard to measure the speech enhancement metrics directly due to the lack of reference signals, we instead evaluate the ASR performance.
The experimental results show that the proposed methods can greatly improve the overall performance of both MC-Conv-TasNet and Beam-TasNet. More than 42\% relative word error rate (WER) reduction is achieved on the evaluation set, while a comparable speech enhancement performance is preserved.

\vspace{-1em}
\section{Proposed Methods}
\label{sec:proposed}
\vspace{-0.7em}
\subsection{Beam-TasNet}
\label{ssec:beam_tasnet}
\vspace{-0.5em}
We first review the Beam-TasNet approach proposed in~\cite{Beam_TasNet-Ochiai2020} and reformulate it in the context of speech enhancement.
The Beam-TasNet system makes use of the MC-Conv-TasNet to estimate speech and noise covariance matrices based on its output signal, and then performs beamforming on the original multi-channel input.

To build this system, the MC-Conv-TasNet model is first trained on simulated multi-channel speech data. As shown in the left part of Figure~\ref{fig:beam_tasnet}, it consists of the multi-channel encoder\footnote{Also called parallel encoder in~\cite{End_to_End-Gu2019}.}, separator, and decoder.
The multi-channel encoder aggregates multiple input channels into one hidden representation, which is then processed by the separator and decoder to generate a single-channel enhanced signal.
In order to generate enhanced signals for all input channels, the MC-Conv-TasNet is trained in a channel-aware manner, which is hereafter referred to as MC-Conv-TasNet*. That is, the original $C$-channel training data is augmented by rotating the input channels anti-clockwise, so that each channel $c$ can be placed as the first channel while preserving the original array geometry. Then, the MC-Conv-TasNet* model is trained to enhance each input signal with the first channel as the reference channel. In the inference phase, the multi-channel output can be obtained by rotating the input channels $C$ times and feeding all channel-rotated~signals~into~the MC-Conv-TasNet*.
The above process can be formulated below:
{\setlength\abovedisplayskip{3pt plus 3pt minus 7pt}
\setlength\belowdisplayskip{3pt plus 3pt minus 7pt}
\begin{align}
    \mathbf{y}^{(c)} &= [y_{c},\ y_{c+1},\ \cdots,\ y_{C}, y_{1},\ y_{2},\ \cdots,\ y_{c-1}]^{\transpose}\,, \label{eq:rotate} \\
    \mathbf{w}_c &= \operatorname{MC-Conv-Encoder}\big(\mathbf{y}^{(c)}\big)\,, \label{eq:mc_enc} \\
    [\mathbf{m}_{c,\text{X}},\ \mathbf{m}_{c,\text{N}}] &= \operatorname{Separator}(\mathbf{w}_c)\,, \label{eq:separator} \\
    \hat{x}_c &= \operatorname{Conv-Decoder}(\mathbf{m}_{c,\text{X}} \otimes \mathbf{w}_c)\,, \label{eq:dec_speech} \\
    \hat{n}_c &= \operatorname{Conv-Decoder}(\mathbf{m}_{c,\text{N}} \otimes \mathbf{w}_c)\,, \label{eq:dec_noise}
\end{align}
}where $\mathbf{y}^{(c)}$ is the channel-rotated input signal with the $c$-th original channel placed at the first channel, $c=1,2,\cdots,C$. $\operatorname{MC-Conv-Encoder}(\cdot)$ is the multi-channel encoder and $\mathbf{w}_c$ is its output representation. $\mathbf{m}_{c, \text{X}}$ and $\mathbf{m}_{c, \text{N}}$ are the predicted speech and noise masks, respectively. $\otimes$ denotes element-wise multiplication. $\hat{x}_c$ and $\hat{n}_c$ are the estimated speech and noise waveforms, respectively.
For training the MC-Conv-TasNet and MC-Conv-TasNet*, we use the combination of time-domain signal-to-noise (SNR) losses on estimated speech and noise signals~\cite{Improving-Kinoshita2020}:
{\setlength\abovedisplayskip{3pt plus 3pt minus 7pt}
\setlength\belowdisplayskip{3pt plus 3pt minus 7pt}
\begin{align}
    \mathcal{L}_{\text{enh}} = -\operatorname{SNR}(x_c, \hat{x}_c) - \operatorname{SNR}(n_c, \hat{n}_c)\,, \label{eq:loss_enh}
\end{align}
}where $\mathbf{x}_c$ and $\mathbf{n}_c$ are the reference speech and noise signals, respectively. $\operatorname{SNR}(x_c, \hat{x}_c)\!\!=\!20 \log_{10}\! \frac{\|x_c\|^2}{\|x_c - \hat{x}_c\|^2}$, and $\|\cdot\|^2$ is the $L_2$ norm.

After training, the Beam-TasNet system is built upon the MC-Conv-TasNet* by calculating the speech and noise covariance matrices based on the enhanced multi-channel speech $\hat{\mathbf{x}}^{(1)} = [\hat{x}_1, \hat{x}_2, \cdots, \hat{x}_C]^{\transpose}$ and then estimating the MVDR filter $\hat{\mathbf{h}}_f$:
{\setlength\abovedisplayskip{5pt plus 3pt minus 7pt}
\setlength\belowdisplayskip{5pt plus 3pt minus 7pt}
\begin{align}
    \hat{\mathbf{h}}_f &= \frac{\big(\mathbf{\Phi}_{\text{N}, f}\big)^{-1} \mathbf{\Phi}_{\text{X}, f}}{\operatorname{Trace}\Big(\big(\mathbf{\Phi}_{\text{N}, f}\big)^{-1} \mathbf{\Phi}_{\text{X}, f}\Big)} \mathbf{u}\,, \label{eq:mvdr}
\end{align}
}where the subscript $f$ is the frequency bin index. $\mathbf{\Phi}_{\text{X},f}$ and $\mathbf{\Phi}_{\text{N},f}$ denote the speech and noise covariance matrices, respectively, which will be discussed in detail in Section~\ref{ssec:integration}. $\operatorname{Trace}(\cdot)$ is the matrix trace operator. $\mathbf{u}$ is a one-hot vector denoting the reference channel $q$.
Then the beamformed signal $\hat{x}_q^{\text{BF}}$ can be derived as follows:
{\setlength\abovedisplayskip{5pt plus 3pt minus 7pt}
\setlength\belowdisplayskip{5pt plus 3pt minus 7pt}
\begin{align}
    \hat{\mathbf{X}}_{f,q}^{\text{BF}} &= \hat{\mathbf{h}}^{\hermconj}\mathbf{Y}_f^{(1)}\,, \label{eq:beamforming} \\
    \hat{x}_q^{\text{BF}} &= \operatorname{iSTFT}\big(\hat{\mathbf{X}}_{q}^{\text{BF}}\big)\,, \label{eq:istft}
\end{align}
}where $\mathbf{Y}_f^{(1)}$ and $\hat{\mathbf{X}}_{f,q}^{\text{BF}}$ are the input noisy spectrum and beamformed spectrum, respectively. $(\cdot)^{\hermconj}$ denotes conjugate transpose. $\operatorname{iSTFT}(\cdot)$ denotes the inverse short-time Fourier transform (STFT).

\vspace{-0.7em}
\subsection{Integration approaches for Beam-TasNet}
\label{ssec:integration}
Following the introduction in Section~\ref{ssec:beam_tasnet} and \cite{Beam_TasNet-Ochiai2020}, there are two main strategies to integrate the MC-Conv-TasNet into the Beam-TasNet architecture, i.e.~sig-MVDR and mask-MVDR.
The sig-MVDR uses the enhanced signals in Eq.~(\ref{eq:dec_speech}) to calculate the covariance matrices in Eq.~(\ref{eq:mvdr}) directly:
{\setlength\abovedisplayskip{3pt plus 3pt minus 7pt}
\setlength\belowdisplayskip{3pt plus 3pt minus 7pt}
\begin{align}
    \mathbf{\Phi}_{\text{X},f} &= \frac{1}{T} \sum_{t} \hat{\mathbf{X}}^{(1)}_{t,f} \big(\hat{\mathbf{X}}^{(1)}_{t,f} \big)^{\hermconj}\,, \label{eq:cov_s1} \\
    \mathbf{\Phi}_{\text{N},f} &= \frac{1}{T} \sum_{t} \big(\mathbf{Y}^{(1)}_{t,f}  - \hat{\mathbf{X}}^{(1)}_{t,f} \big) \big(\mathbf{Y}^{(1)}_{t,f}  - \hat{\mathbf{X}}^{(1)}_{t,f} \big)^{\hermconj}\,, \label{eq:cov_n1}
\end{align}
}where $\hat{\mathbf{X}}^{(1)}_{t,f}$ is the speech spectrum enhanced by MC-Conv-TasNet. The mask-MVDR estimates time-frequency masks from the enhanced signals for calculating the covariance matrices:
{\setlength\abovedisplayskip{3pt plus 3pt minus 7pt}
\setlength\belowdisplayskip{3pt plus 3pt minus 7pt}
\begin{align}
    \mathbf{\Phi}_{\alpha,f} &= \frac{1}{T} \sum_{t} M_{\alpha,t,f} \mathbf{Y}^{(1)}_{t,f} \big(\mathbf{Y}^{(1)}_{t,f} \big)^{\hermconj}\,, \label{eq:cov_s2}
\end{align}
}where $\alpha \in \{\text{X}, \text{N}\}$. $M_{\alpha,t,f}$ is the estimated speech\,/\,noise mask.

In this section, we would like to put more emphasis on the mask-MVDR strategy, as the sig-MVDR based method may unexpectedly corrupt the appropriate spatial correlation in the multi-channel signal, while the mask-MVDR based method can mitigate such corruption, which will be shown later in our experiments.
Since the speech mask is estimated from the enhanced signal, various types of masks can be investigated, including the well-known phase-sensitive mask (PSM)~\cite{Phase_sensitive-Erdogan2015}, and the voice activity detection (VAD) like 1-D mask.
The PSM takes into account the phase information explicitly, which can be beneficial to the covariance matrix estimation.
The VAD-like 1-D mask is shown to be robust against noise or interference signals~\cite{Iterative-Tu2019, Investigation-Subramanian2019, End_to_End-Zhang2021}, and it is calculated by averaging the power mask $\mathbf{M}_{\text{X}}$ along the frequency dimension:
{\setlength\abovedisplayskip{5pt plus 3pt minus 7pt}
\setlength\belowdisplayskip{5pt plus 3pt minus 7pt}
\begin{align}
    M_{\text{X},t}^{\text{1-D}} = \frac{1}{F}\sum_{f} M_{\text{X},t,f}\,. \label{eq:vad_mask}
\end{align}
}

\vspace{-2em}
\subsection{Joint training of MC-Conv-TasNet and ASR}
\label{ssec:joint_train}
\vspace{-.25em}
Another direction is to jointly optimize the MC-Conv-TasNet frontend and the end-to-end ASR backend, which implicitly mitigates the mismatch between speech enhancement and ASR systems.

Since MC-Conv-TasNet directly operates on the raw waveform, the memory consumption can be very large when processing a full-length waveform, making it impractical for joint training. In order to jointly optimize MC-Conv-TasNet and end-to-end (E2E) ASR, we adopt the approximated truncated back-propagation through time (TBPTT) strategy used in~\cite{End_to_end-vonNeumann2020}.
That is, the backward graph is only retained for a randomly selected chunk instead of the full-length waveform, while the other part of the waveform is used only for the forward pass. Then, the full-length enhanced signal with the partially retained backward graph is fed into the ASR backend.
This enables us to jointly train both frontend and backend with a flexibly adjustable memory cost, which is determined by the chunk size $K$.

When simulated and real data with transcripts are available for training, our proposed framework allows exploiting both data for optimizing the SE and ASR models. As illustrated in Figure~\ref{fig:joint}, the final loss $\mathcal{L}$ in the joint training is composed of two parts, i.e.~the speech enhancement $\mathcal{L}_{\text{enh}}$ and the ASR loss $\mathcal{L}_{\text{asr}}$.
For real data, the final loss is equal to $\mathcal{L}_{\text{asr}}$, i.e.~we train the SE and ASR models end-to-end without the need of signal-level references.
For simulated data, the final loss is defined as $\mathcal{L} = \mathcal{L}_{\text{enh}} + \mathcal{L}_{\text{asr}}$.

Furthermore, the clean speech data can be additionally utilized to train only the ASR backend.
The above multi-condition training strategy provides a flexible way for the MC-Conv-TasNet model to adapt to both simulated and real data, which is shown to greatly improve the ASR performance on real evaluation data in Section~\ref{ssec:exp_chime4}.

\begin{figure}[t]
  \centering
  \centerline{\includegraphics[width=0.7\columnwidth]{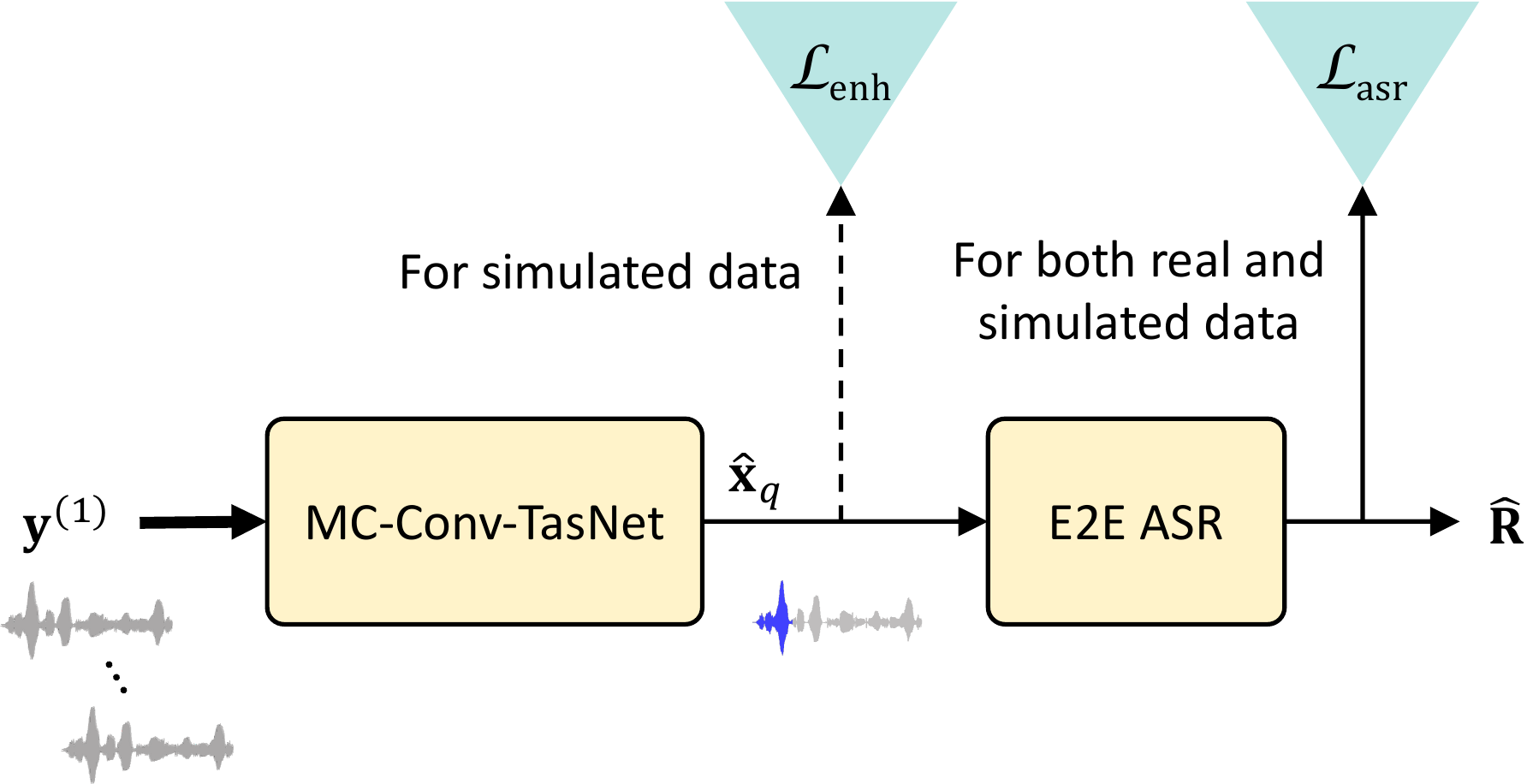}}
  \vspace{-1em}
  \caption{Joint training of SE and ASR models with truncated back-propagation through time. The blue and gray chunks denote the enhanced signal with and without the backward graph, respectively.}
  \label{fig:joint}
  \vspace{-.5em}
\end{figure}
\vspace{-1.3em}
\section{Experiments}
\label{sec:exp}
\vspace{-0.8em}
\subsection{Experimental Setup}
\label{ssec:setup}
\vspace{-0.3em}
We conducted experiments on the 6-channel track of the CHiME-4~\cite{CHiME4-Vincent2017} corpus to evaluate our proposed methods.
The CHiME-4 corpus consists of both real recordings and simulated speech data, so that we can easily evaluate the robustness and generalizability of our proposed approaches in unseen conditions.
There are 42828 (9600), 1640 (1640), and 1320 (1320) simulated (real) samples for training, development and evaluation, respectively.
The sample rate is 16 kHz for all speech data.
We adopt the 5-th channel (CH5) as the reference channel for both training and evaluation.
For evaluating the performance of frontend models on the real data, we adopt an E2E ASR model pretrained on the CHiME-4 dataset, which was also used in Section 4.1 in~\cite{ESPnet_SE-Li2021}.
For the joint training of frontend and backend, we optionally include an additional dataset from the Wall Street Journal (WSJ) corpus~\cite{wsj0} for training, which consists of 37416 clean speech samples.
SpecAugment~\cite{SpecAugment-Park2019} is applied to the ASR input feature during the joint training.
The chunk size $K$ mentioned in Section~\ref{ssec:joint_train} is set to 3 seconds.
We use the Adam optimizer for model training, and all models are trained until convergence.

All our models are built based on the ESPnet toolkit~\cite{ESPnet-Watanabe2018,ESPnet_SE-Li2021}. The MC-Conv-TasNet model uses a Conv1D layer with 5 input channels and 256 output channels for the multi-channel encoder, with a kernel size of 20 and stride of 10.
The separator consists of 4 stacked dilated convolutional blocks, each composed of 8 Conv1D blocks with 256 bottleneck channels and 512 hidden channels.
The decoder is a transposed Conv1D layer with 256 input channels and 1 output channel, and the kernel size and stride are the same as the multi-channel encoder.
The E2E ASR model is a joint connectionist temporal classification (CTC)/attention-based encoder-decoder~\cite{Joint-Kim2017} model, which consists of 12 and 6 transformer blocks with 2048 hidden units and four 64-dimensional attention heads for the encoder and decoder, respectively.

For speech enhancement performance, we adopt the short-time objective intelligibility (STOI) \cite{Algorithm-Taal2011}, perceptual evaluation of speech quality (PESQ) \cite{Perceptual-Rix2001}, and signal-to-distortion ratio (SDR) for evaluation. For ASR performance, the WER is evaluated.

\begin{table*}
    \caption{Performance (PESQ\,/\,STOI\,/\,SDR [dB]\,/\,WER [\%]) on the CHiME-4 6-channel track. The same ASR model pretrained on CHiME-4 is used for evaluating WER on models No.~2{\around}No.~10. For PESQ, STOI, and SDR, larger is better. For WER (gray cells), smaller is better.}
    \label{tab:exp_enh}
    \centering
    \resizebox{\textwidth}{!}{%
    \begin{tabular}{c l | c c c c | c c c c | c | c}
        \toprule
        \multirow{2}{*}{\textbf{No.}} & \multirow{2}{*}{\textbf{Model}} & \multicolumn{4}{c|}{\textbf{Dev (Simu)}} & \multicolumn{4}{c|}{\textbf{Test (Simu)}} & \textbf{Dev (Real)} & \textbf{Test (Real)} \\
        & & \textbf{PESQ} & \textbf{STOI} & \textbf{SDR} & \textbf{WER} & \textbf{PESQ} & \textbf{STOI} & \textbf{SDR} & \textbf{WER} & \textbf{WER} & \textbf{WER} \\
        \midrule
        1 & Official baseline in~\cite{CHiME4-Vincent2017} & - & - & - & \cellcolor[HTML]{E0E0E0}6.8 & - & - & - & \cellcolor[HTML]{E0E0E0}10.9 & \cellcolor[HTML]{E0E0E0}5.8 & \cellcolor[HTML]{E0E0E0}11.5 \\
        \midrule
        2 & Noisy Input (CH5) & 2.17 & 0.86 & 5.78 & \cellcolor[HTML]{E0E0E0}12.6 & 2.18 & 0.87 & 7.54 & \cellcolor[HTML]{E0E0E0}19.9 & \cellcolor[HTML]{E0E0E0}10.9 & \cellcolor[HTML]{E0E0E0}19.5 \\
        3 & BeamformIt in~\cite{ESPnet_SE-Li2021} & 2.31 & 0.88 & 5.51 & \cellcolor[HTML]{E0E0E0}8.4 & 2.20 & 0.86 & 6.25 & \cellcolor[HTML]{E0E0E0}13.9 & \cellcolor[HTML]{E0E0E0}7.3 & \cellcolor[HTML]{E0E0E0}13.2 \\
        4 & BLSTM MVDR in~\cite{ESPnet_SE-Li2021} & 2.68 & 0.95 & 13.40 & \cellcolor[HTML]{E0E0E0}\textbf{5.3} & 2.68 & 0.95 & 14.10 & \cellcolor[HTML]{E0E0E0}8.0 & \cellcolor[HTML]{E0E0E0}\textbf{5.9} & \cellcolor[HTML]{E0E0E0}\textbf{9.8} \\
        5 & FaSNet~\cite{FaSNet-Luo2019} & 2.64 & 0.93 & 10.56 & \cellcolor[HTML]{E0E0E0}8.3 & 2.43 & 0.89 & 9.73 & \cellcolor[HTML]{E0E0E0}18.4 & \cellcolor[HTML]{E0E0E0}10.3 & \cellcolor[HTML]{E0E0E0}22.5 \\
        \midrule
        6 & MC-Conv-TasNet & 3.08 & 0.96 & 18.32 & \cellcolor[HTML]{E0E0E0}6.2 & 2.92 & 0.95 & 17.52 & \cellcolor[HTML]{E0E0E0}10.4 & \cellcolor[HTML]{E0E0E0}22.1 & \cellcolor[HTML]{E0E0E0}33.2 \\
        7 & MC-Conv-TasNet* & \textbf{3.08} & \textbf{0.97} & \textbf{18.62} & \cellcolor[HTML]{E0E0E0}6.7 & 2.90 & 0.95 & \textbf{17.99} & \cellcolor[HTML]{E0E0E0}11.4 & \cellcolor[HTML]{E0E0E0}18.5 & \cellcolor[HTML]{E0E0E0}41.5 \\
        8 & $\ \ \rightarrow$ Beam-TasNet (sig-MVDR) & 2.57 & 0.95 & 15.31 & \cellcolor[HTML]{E0E0E0}\textbf{5.3} & 2.58 & 0.95 & 16.17 & \cellcolor[HTML]{E0E0E0}\textbf{7.3} & \cellcolor[HTML]{E0E0E0}11.8 & \cellcolor[HTML]{E0E0E0}25.7 \\
        9 & $\ \ \rightarrow$ Beam-TasNet (mask-MVDR, PSM) & 2.61 & 0.95 & 14.78 & \cellcolor[HTML]{E0E0E0}5.4 & 2.65 & 0.95 & 15.25 & \cellcolor[HTML]{E0E0E0}7.4 & \cellcolor[HTML]{E0E0E0}7.2 & \cellcolor[HTML]{E0E0E0}13.5 \\
        10 & $\ \ \rightarrow$ Beam-TasNet (mask-MVDR, 1-D) & 2.62 & 0.95 & 14.11 & \cellcolor[HTML]{E0E0E0}5.7 & 2.66 & 0.95 & 15.94 & \cellcolor[HTML]{E0E0E0}7.7 & \cellcolor[HTML]{E0E0E0}6.3 & \cellcolor[HTML]{E0E0E0}10.7 \\
        \midrule
        11 & Joint MC-Conv-TasNet + ASR & 3.06 & \textbf{0.97} & 18.27 & \cellcolor[HTML]{E0E0E0}9.3 & \textbf{2.96} & \textbf{0.96} & 17.52 & \cellcolor[HTML]{E0E0E0}13.7 & \cellcolor[HTML]{E0E0E0}19.0 & \cellcolor[HTML]{E0E0E0}42.5 \\
        12 & $\ \ $+ real training data & 3.07 & 0.96 & 18.19 & \cellcolor[HTML]{E0E0E0}8.3 & 2.93 & 0.95 & 17.39 & \cellcolor[HTML]{E0E0E0}12.1 & \cellcolor[HTML]{E0E0E0}9.1 & \cellcolor[HTML]{E0E0E0}17.0 \\
        13 & $\quad$++ clean training data & 3.05 & 0.96 & 18.10 & \cellcolor[HTML]{E0E0E0}6.5 & 2.93 & 0.95 & 17.24 & \cellcolor[HTML]{E0E0E0}10.0 & \cellcolor[HTML]{E0E0E0}7.3 & \cellcolor[HTML]{E0E0E0}13.5 \\
        \bottomrule
    \end{tabular}%
    }
    \vspace{-0.5em}
\end{table*}
\begin{figure*}
  \centering
  \centerline{\includegraphics[width=\textwidth]{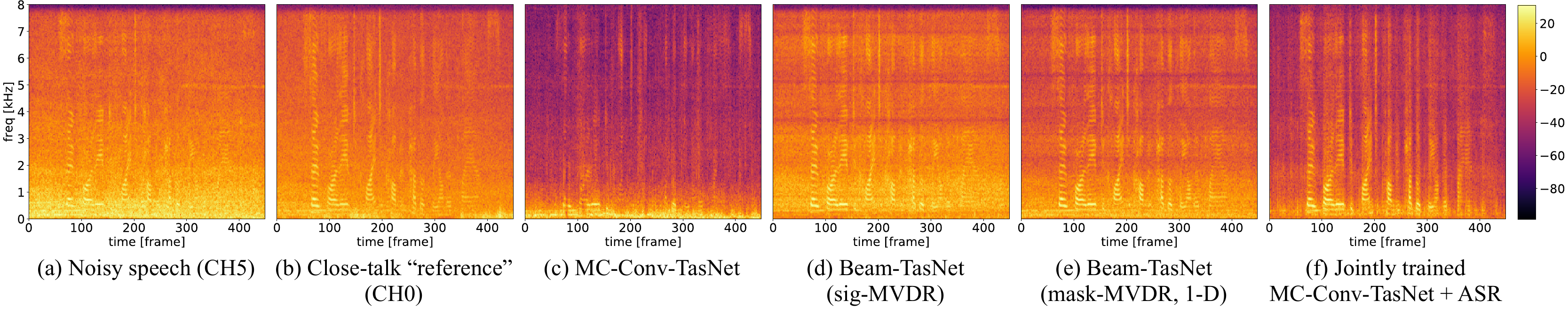}}
  \vspace{-1em}
  \caption{Spectrograms of a real speech recording randomly selected from the CHiME-4 evaluation set.}
  \label{fig:spec}
  \vspace{-1.4em}
\end{figure*}
\vspace{-1.4em}
\subsection{Performance evaluation on simulation and real data}
\label{ssec:exp_chime4}
\vspace{-.6em}
The performance of the baseline methods (No.~1{\around}No.~5) and proposed methods (No.~6{\around}No.~13) are listed in Table~\ref{tab:exp_enh}.
Here we take the official result~\cite{CHiME4-Vincent2017} as the ASR baseline, which uses a DNN-HMM acoustic model with language model rescoring. The speech enhancement baselines include the BeamformIt~\cite{Acoustic-Anguera2007}, neural beamformer (denoted as ``BLSTM MVDR'') from~\cite{ESPnet_SE-Li2021}, and the time-domain filter-and-sum network (FaSNet)\footnote{We use the open-source implementation at \url{https://github.com/yluo42/TAC/blob/master/FaSNet.py\#L176}.}~\cite{FaSNet-Luo2019}.
The speech recognition performance is evaluated using the same pretrained E2E ASR model on CHiME-4 for models from No.~2 to No.~10.

For the speech enhancement performance on simulated data, it can be observed that all MC-Conv-TasNet based models outperform the baselines, and the best performance is achieved by the MC-Conv-TasNet* model, which is trained using the channel-rotated data as described in Section~\ref{ssec:beam_tasnet}.
Compared to No.~2, the WERs of the MC-Conv-TasNet models on simulated data are also greatly reduced.
However, the WERs on the real data become worse than the ASR baseline, which indicates the over-training of the MC-Conv-TasNet models in simulation conditions.
In contrast, the frequency-domain BLSTM MVDR model shows better generalizability.\footnote{Another way to define the performance gap is the relative WER ratio between Real and Simu conditions, and we observed similar conclusions with this metric.}

After applying the trained MC-Conv-TasNet* model to the Beam-TasNet framework (No.~8{\around}No.~10), we can observe significant WER reduction on both development and evaluation sets, especially on the real data. This is attributed to the distortionless constraint that is explicitly enforced in the design of MVDR beamforming. On the other hand, the speech enhancement performance of Beam-TasNet is worse than the MC-Conv-TasNet*, which could result from the fact that MVDR beamforming does not fully eliminate the noise and the residual noise level may be higher than that in MC-Conv-TasNet*.
In addition, we can observe that the mask-MVDR based Beam-TasNet achieves much better speech recognition performance on real data than the sig-MVDR based one, while sacrificing some speech enhancement performance. This illustrates that the masking based integration approach can effectively mitigate the artifacts introduced in the TasNet output, making it more practical in realistic scenarios. And the VAD-like 1-D mask shows better ASR performance than PSM, which can be attributed to the averaging operation in Eq.~(\ref{eq:vad_mask}) that further eliminates some inaccurate estimations.
Compared to the baseline ASR model, more than 42\% relative WER reduction is achieved on all subsets.

Finally, the joint training\footnote{We jointly finetuned the pretrained MC-Conv-TasNet and E2E ASR models instead of training from scratch.} in the last three lines leads to a comparable speech enhancement performance to the MC-Conv-TasNet models.
Although the ASR performance on the simulated data is slightly worse than the MC-Conv-TasNet result (No.~6), this can be regarded as the effect of regularization from the real data during training.
When only the simulated data is used for training (No.~11), the speech recognition on real data is severely degraded compared to the ASR baseline, which indicates the over-training in the simulation condition.
After introducing the real data (No.~12) and additionally the clean data (No.~13) for training, we can observe pronounced WER reduction on the real data. This illustrates the advantage of joint training that various training data from different conditions can be well utilized to improve the overall performance of the entire system.
Although the ASR performance does not outperform the BLSTM MVDR model, the performance gap is largely reduced, and much better enhancement performance is achieved.

To better illustrate the discrepancy between different speech enhancement methods, we further visualize the original spectrogram of a real sample and its enhanced versions from different models in Figure~\ref{fig:spec}. Subfigures (a) and (b) are the noisy speech recorded by the distant microphone and close-talk microphone, respectively. Subfigures (c)--(f) show the corresponding enhanced signals by different models discussed above.
We can observe that the MC-Conv-TasNet model severely corrupts the speech pattern in the spectrum, while the Beam-TasNet and jointly trained MC-Conv-TasNet models can restore the speech pattern and suppress the noise to some extent. This observation also coincides with the results in Table~\ref{tab:exp_enh}.\footnote{We also conducted subjective evaluation and the result is available at \url{https://x-lance.sjtu.edu.cn/members/wangyou-zhang/waspaa21_subjective_evaluation}.}

\vspace{-1.15em}
\section{Conclusion}
\label{sec:conclusion}
\vspace{-.9em}
In this paper, we investigate the performance of multi-channel Conv-TasNet based models for time-domain speech enhancement. A large performance gap is observed between simulation and real conditions. And several approaches are proposed to reduce this gap and improve the robustness of MC-Conv-TasNet based models, including the integration into the Beam-TasNet framework, and the joint training of MC-Conv-TasNet and ASR models.
Experimental results on the CHiME-4 data show the difficulty of achieving good performance on real data, and that well-trained speech enhancement models on the simulated data do not necessarily remain advantageous when evaluated on real data.
Our proposed approaches are shown to effectively mitigate the ASR performance gap, while still preserving a comparable speech enhancement capability.
In the future work, we would like to incorporate more approaches to mitigating the mismatch between real and simulation conditions, such as better simulation strategies.

\vspace{-1.1em}
\section{ACKNOWLEDGMENT}
\label{sec:ack}
\vspace{-0.85em}
The authors would like to thank Dr.~Tsubasa Ochiai for his helpful comments about the gap between Beam-TasNet enhancement results on real and simulation conditions.
This work was supported by the China NSFC projects (No. 62071288 and No. U1736202) and Shanghai Municipal Science and Technology Major Project (2021SHZDZX0102). Experiments have been carried out on the PI super-computer at Shanghai Jiao Tong University.

\bibliographystyle{IEEEtran}
\bibliography{refs21}

\end{sloppy}
\end{document}